\documentstyle[11pt,newpasp,twoside,epsf]{article}
\def\edcomment#1{\iffalse\marginpar{\raggedright\sl#1\/}\else\relax\fi}
\reversemarginpar

\begin{document}
\title{Molecular emission from the shocked bipolar outflow in OH\,231.8+4.2}
\author{C. S\'anchez Contreras$^{1,2}$, V. Bujarrabal$^{1}$, R. Neri$^{3}$, 
J. Alcolea$^{1}$}
\affil{$^{1}$ Observatorio Astron\'omico Nacional, Ap. 1143, 
       E-28800 Alcal\'a de Henares, Spain}
\affil{$^{2}$ Dpto$.$ de Astrof\'{\i}sica, F.CC$.$ F\'{\i}sicas, 
       UCM, E-28040 Madrid, Spain}
%\author{R. Neri}
\affil{$^{3}$ IRAM, 300 rue de la Piscine, F-38406 St Martin d'H\`eres, France}

\begin{abstract}
       We present high-resolution observations of several molecular
       lines in OH\,231.8+4.2 taken with the IRAM interferometer. All
       molecules are distributed in a narrow region along the symmetry
       axis, and flow outwards following a velocity gradient similar
       to that found in CO. The HCO$^{+}$ emission is found to be very
       clumpy and strongly enhanced in the shock-accelerated lobes,
       indicating that the formation of this molecule is probably
       dominated by shock induced reactions. SO is present in the
       axial outflow as well as in an expanding equatorial disk. The
       SiO maser emission seems to arise from the innermost parts of
       such a disk. We also report the first detection of NS in
       circumstellar envelopes.

\end{abstract}

%\vspace{-0.75cm}
\section{Introduction}
        OH\,231.8+4.2 (hereafter OH\,231.8) is a remarkable bipolar
        nebula located \linebreak $\sim$\,1500\,pc away (Kastner et
        al$.$ 1992, Bowers \& Morris 1984) that surrounds a cold
        (M\,9\,III) Mira star (Cohen 1981; Kastner et al$.$ 1998). The
        optical nebula consists of two extended lobes, oriented at
        position angle $\sim$ 21\deg\ and inclined with respect to the
        plane of the sky about 36\deg\ (Kastner et al$.$ 1992). The
        gas in the lobes is expanding at high velocity from the nebula
        center and its optical emission-line spectrum indicates that
        it has been excited by shocks (Cohen et al$.$ 1985; S\'anchez
        Contreras et al$.$ 1999). The molecular envelope of OH\,231.8
        is very cold (T$_{\rm kin}\sim$\,10\,K) and massive ($\sim$
        1$M_{\sun}$). This envelope is expanding at low velocity
        ($\sim$\,10--15\,km\,s$^{-1}$) near the equator, while at
        higher latitudes a strong axial expansion appears. In contrast
        to the atomic material, the molecular gas is highly restricted
        to the symmetry axis of the nebula (see Alcolea et al$.$ in
        this volume). The pronounced axial symmetry and the large
        velocities reached by the gas in OH\,231.8 are usually
        attributed to the impact of a recent, highly collimated wind
        on the old AGB envelope.

        A large variety of molecules has been detected in
        OH\,231.8 (e$.$g$.$ Morris et al$.$ 1987),
        which is classified as an O-rich source due to its H$_{2}$O,
        OH and SiO maser emission. Maps of certain molecules
        have suggested that an active chemistry, probably
        induced by shocks, takes place in OH\,231.8 (S\'anchez 
	Contreras et al$.$ 1997; Jackson \& Nguyen-Q-Rieu 1988).

\section{Observations and results}
	We have obtained high-resolution maps of the
	HCO$^{+}$\,($J$=1--0), SO\,($J$=2$_{2}$--1$_{1}$),
	H$^{13}$CN\,($J$=1--0), SiO\,($v$=1, $J$=2--1), and NS
	($^{2}\Pi_{1/2}$, $J$=5/2--3/2, parity-$e$) lines in
	OH\,231.8. Observations were performed using the IRAM
	interferometer (Plateau de Bure, France) during the winters of
	96-97 and 97-98, and in February 99. The highest spatial and
	spectral resolution obtained are $\sim$\,3\arcsec\ and
	$\sim$0.3\,km\,s$^{-1}$, respectively.  The spatial origin of
	our maps coincides with the position of the maser, that is
	located at R.A.= 07$^{\rm h}$42$^{\rm m}$16\fs93,
	Dec.=$-$14\deg\,42\arcmin50\farcs2 (J2000).

%%%%%%%%%%%%%%%%%%%%%%%%%%%%%%%%%%%%%%%%%%%%%%%%%%%%%%%%%%%%%%%%%%%%%%%%%%%
\begin{figure}
\plotfiddle{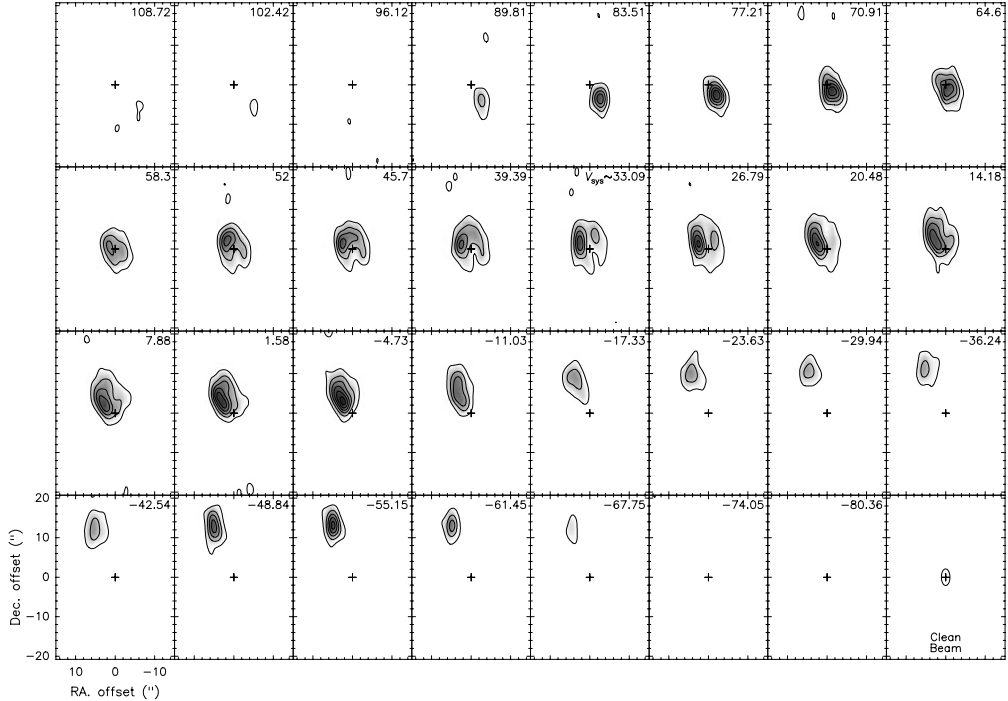}{8.25cm}{270}{52}{52}{-220}{285}
\caption{HCO$^{+}$ maps for channels at different LSR velocities 
(top-right corners). Levels are 0.01 to 0.11 by 0.02 Jy/beam}
\end{figure}
%%%%%%%%%%%%%%%%%%%%%%%%%%%%%%%%%%%%%%%%%%%%%%%%%%%%%%%%%%%%%%%%%%%%%%%%%%%

\subsection{HCO$^{+}$}
	In Fig$.$\,1 we present the interferometric maps of the
	HCO$^{+}$(1--0) line for different velocity channels. The
	emission is distributed in a narrow region along the symmetry
	axis of the nebula. Such a region has a total size of about
	(1$\times$7)\,10$^{17}$\,cm, and seems to be a hollow cylinder
	near the equator. The present high-resolution maps show that
	the HCO$^{+}$ emission is notably clumpy. We confirm the axial
	velocity gradient (Fig$.$\,2, top-left box) found in previous
	low-resolution observations. This gradient is almost constant
	along the axis and similar to that found for $^{12}$CO and
	other molecules (see below and S\'anchez Contreras et al$.$
	1997). The position-velocity (p-v) diagram along the nebula
	equator (Fig$.$\,2, top-right box) roughly corresponds to an
	expanding, hollow cylinder or ring. It is remarkable that the
	most intense HCO$^{+}$ emission arises from regions moving at
	high velocities. In fact, there is a local minimum in the
	emission of the low-velocity expanding component (between
	10--55\,km\,s$^{-1}$) near the center, where the rest of the
	observed molecules reach the maximum intensity (see SO data in
	Fig$.$\,2 and CO in Alcolea et al$.$ this volume).  The
	observed spatial and spectral distributions of HCO$^{+}$
	suggest that this molecule is efficiently formed in the
	accelerated gas by shock-induced reactions.

%%%%%%%%%%%%%%%%%%%%%%%%%%%%%%%%%%%%%%%%%%%%%%%%%%%%%%%%%%%%%%%%%%%%%%%%%%%
\begin{figure}
\plotfiddle{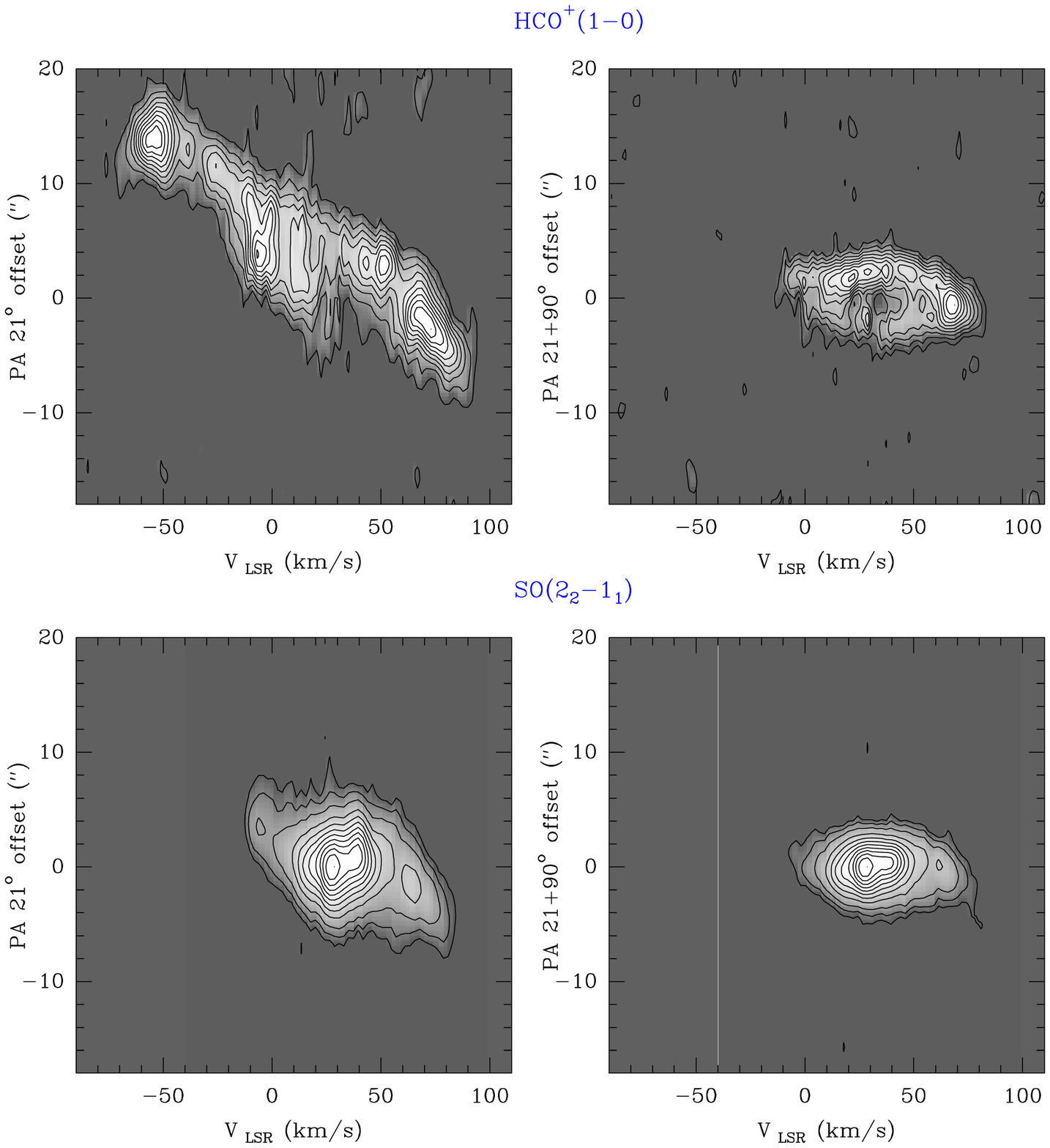}{7.2cm}{0}{40}{40}{-205}{-23.5}
\plotfiddle{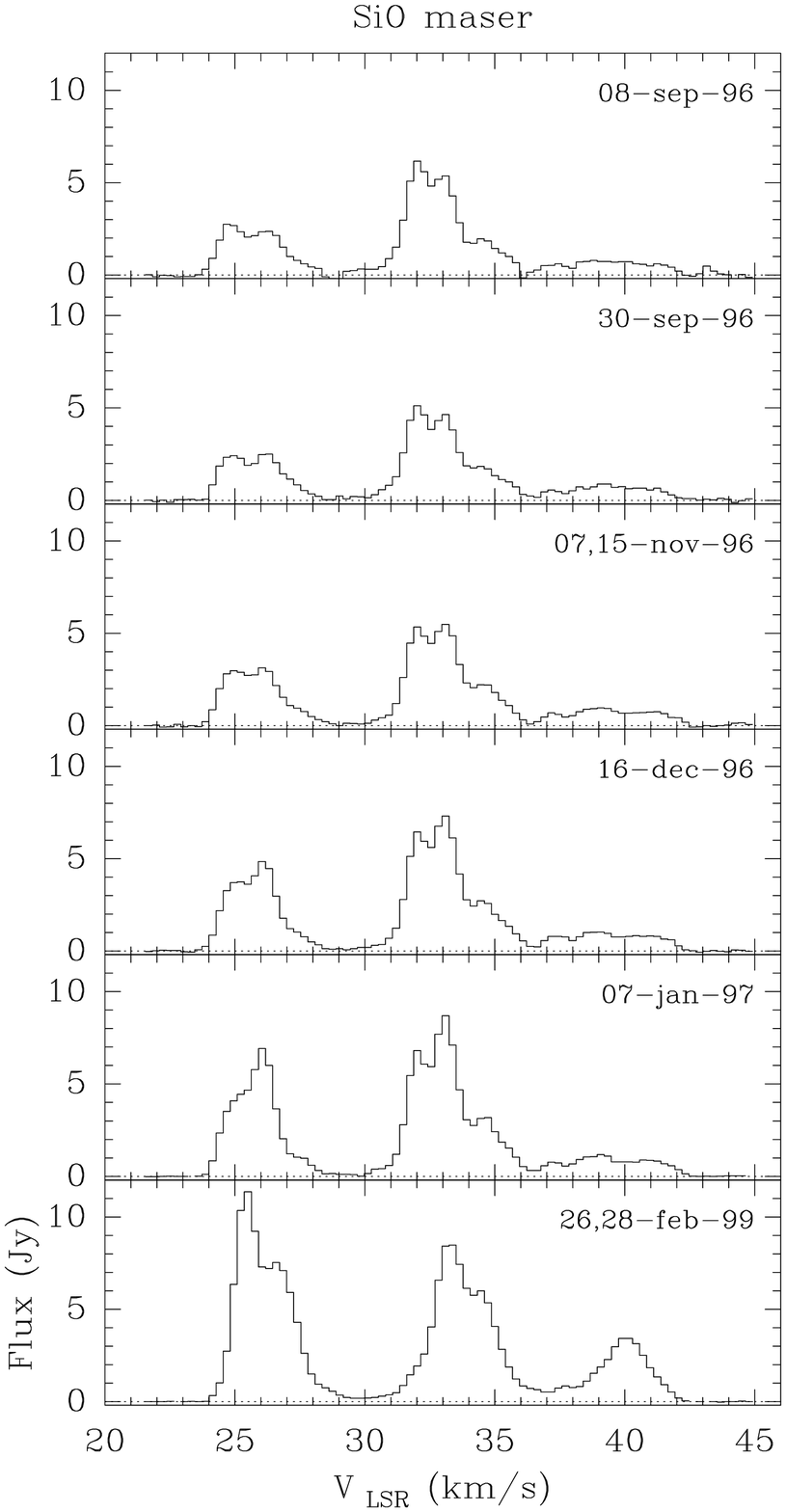}{0.01cm}{0}{30}{30}{55}{36}
\hspace{-6cm} \parbox[b]{8.4cm}{\caption[]{HCO$^{+}$ and SO p-v diagrams. 
Levels for HCO$^{+}$ (SO) are 10 (2.5, 5, 10) to 100 by 10 
\% of the maximum}}

\vspace{-1.9cm} \hspace{6.6cm} \parbox[t]{6.2cm}{\caption[]{SiO maser 
spectra in different epochs}} \vspace{0.4cm}
\end{figure}
%%%%%%%%%%%%%%%%%%%%%%%%%%%%%%%%%%%%%%%%%%%%%%%%%%%%%%%%%%%%%%%%%%%%%%%%%%%

\subsection{SO}
	The SO emission occupies a narrow region
	($\sim$\,10$^{17}$\,cm) extending
	$\sim$\,3.5\,10$^{17}$\,cm along the symmetry axis of the
	nebula.  The total line width is $\sim$\,100\,km\,s$^{-1}$,
	indicating that SO is present in the accelerated lobes of
	OH\,231.8. In Fig$.$\,2 (bottom-left box) we can see that SO
	follows the general velocity gradient along the symmetry axis
	of the nebula. We have also found SO emission in an expanding
	disk surrounding the central star. The presence of such a disk
	is indicated by the inversion of the slope of the velocity
	gradient at the nebula center (Fig$.$\,2, bottom-left
	box). The characteristic radius and the expansion velocity of the
	equatorial disk are $\sim$\,2\,10$^{16}$\,cm and
	$\sim$\,7\,km\,s$^{-1}$, respectively. These values lead to
	a kinematical age for the disk of $\sim$\,1000\,yr, very
	similar to that found for the bipolar molecular flow. No
	sign of rotation has been found in the disk (Fig$.$\,2).
	Differences between the SO abundaces in the disk and the outflow 
	are negligible.

\subsection{The SiO maser}
	In Fig$.$\,3 we show the SiO ($v$=1, $J$=2--1) maser spectra
	in six different epochs. Three main spectral components can be
	distinguished at (LSR) velocities $\sim$\,26, $\sim$\,33 (the
	systemic velocity), and $\sim$\,40\,km\,s$^{-1}$. Each of
	these features is formed of (at least) 3 components,
	indicating the complex and, probably, clumpy distribution of
	the gas in the vicinity of the central star. We estimate from
	our mapping that the size of the maser emitting region is
	$\la$\,3\,10$^{15}$\,cm.  The fact that the three main
	spectral features are also found in the SO line (Fig$.$\,2)
	suggests that the maser emission arises from the innermost
	regions of the expanding disk seen in the SO maps. Our data
	would indicate a negligible velocity gradient between the
	inner and outer parts of such a disk. The relative intensity
	between the different spectral components as well as the total
	flux of the SiO line is found to strongly vary with time
	(Fig$.$\,3).  We have found a relative minimum of the SiO flux
	$\sim$\,60 days before the n-IR minimum (Kastner et al$.$
	1992).  Assuming that maser pumping is radiative, this phase
	lag would correspond to a distance of $\sim$\,10$^{17}$\,cm
	between the maser (very close to the star) and the nebular
	dust reflecting the n-IR starlight. This value is in agreement
	with the measured distance from the center to the region
	with maximum n-IR emission.

\subsection{NS and H$^{13}$CN}
	We report the first detection of nitrogen sulfide
	($^{2}\Pi_{1/2}$, $J$=5/2--3/2, parity-$e$) in circumstellar
	envelopes. The NS emission is found in a compact central
	region and, tentatively, in the outflow. The total flux of the
	line (integrated over all the hyperfine components) is
	relatively high, $\sim$ 15\,Jy\,km\,s$^{-1}$.

	We have found the H$^{13}$CN\,($J$=1--0) emission to be
	distributed along the molecular outflow sharing the general
	velocity gradient.  From the H$^{12}$CN/H$^{13}$CN ($J$=1--0)
	intensity ratio (H$^{12}$CN data from S\'anchez Contreras et
	al$.$ 1997) we deduce an isotopic $^{12}$C/$^{13}$C ratio of
	$\sim$\,5--10.

	The large variety and abundance of N- and S-bearing molecules
	in OH\,231.8 is a clear sign of an active chemistry probably
	induced by shocks. Shocks would iniciate (endothermic)
	reactions that trigger the N and S chemistry (Lada,
	Oppenheimer \& Hartquist 1978), and could also extract
	additional S from the surface of dust grains (Jackson \&
	Nguyen-Q-Rieu 1988).

\acknowledgments
C.S.C. thanks the Organizing Committee of
this conference for a registration fee waiver. This work has been 
partially supported by DGES project number PB96-0104.

\end{document}